\documentclass[12pt]{article}
\newcommand{\ba}{\begin{array}}
\newcommand{\ea}{\end{array}}

\newcommand{\be}{\begin{equation}}
\newcommand{\ee}{\end{equation}}
\newcommand{\bea}{\begin{eqnarray}}
\newcommand{\eea}{\end{eqnarray}}
\newcommand{\beal}{\setcounter{letter}{1} \begin{eqnarray}}
\newcommand{\eeal}{\addtocounter{equation}{1} \end{eqnarray}}
\newcommand{\none}{\nonumber \\}

\newcommand{\larrow}{\,\,\,\,\hbox to 30pt{\rightarrowfill}
\,\,\,\,}
\newcommand{\slarrow}{\,\,\,\hbox to 20pt{\rightarrowfill}
\,\,\,}

\newcommand{\bm}{\bibitem}
\begin{document}

\begin{titlepage}
\renewcommand{\thefootnote}{\fnsymbol{footnote}}
\renewcommand{\baselinestretch}{1.3}
\medskip
\hfill  UNB Technical Report 05-03\\[20pt]

\begin{center}
{\large {\bf The Generalized Ricci Flow for 3D Manifolds with One Killing Vector }}
\\ \medskip  {}
\medskip

\renewcommand{\baselinestretch}{1}
{\bf
J. Gegenberg $\dagger$
G. Kunstatter $\sharp$
\\}
\vspace*{0.50cm}
{\sl
$\dagger$ Dept. of Mathematics and Statistics and Department of Physics,
University of New Brunswick\\
Fredericton, New Brunswick, Canada  E3B 5A3\\

}
{\sl
$\sharp$ Dept. of Physics and Winnipeg Institute of
Theoretical Physics, University of Winnipeg\\
Winnipeg, Manitoba, Canada R3B 2E9\\

}

\end{center}

\renewcommand{\baselinestretch}{1}

\begin{center}
{\bf Abstract}
\end{center}
{\small
We consider  3D flow equations inspired by the renormalization group (RG) equations of string theory with a three dimensional target space. By modifying the flow equations to include a U(1) gauge field, and adding carefully chosen De Turck terms, we are able to extend recent 2D results of Bakas to the case of a 3D  Riemannian metric with one Killing vector. In particular, we show that the RG flow 
with De Turck terms can be reduced to two equations:  the continual Toda flow 
solved by Bakas, plus its linearizaton. We find exact solutions which flow to homogeneous but not always isotropic geometries. }
\vfill
\hfill  September 2005 \\
\end{titlepage}

\section{Introduction}
The Ricci flow of d-dimensional manifolds is interesting because of its relationship
to the renormalization group equations of generalized 2D sigma models with d-dimensional target space. In two spacetime dimensions, Ricci flow also provides a proof of the uniformization theorem\cite{poincare}, which states
that every closed orientable
two dimensional manifold with handle number 0,1, or $>1$ admits
{\it uniquely}
the constant curvature geometry with positive, zero,
or negative curvatures, respectively.  
Bakas \cite{bakas} has shown that the 2D Ricci flow equations in conformal gauge
provide a continual analogue of the Toda field equations. Using this algebraic approach
 he was able to write down the general solution.

The potential importance of a 3D uniformization theorem
is evident particularly in the context of
(super)membrane physics and three-dimensional quantum gravity
where one should be able to perform path-integral quantization via a similar
procedure to that in two dimensions.
Unfortunately, there is no uniformization theorem in three dimensions, only
a conjecture due to W.P. Thurston.
\cite{thurston,scott}.  

Recently there has been speculation that Perelman
\cite{perelman} has overcome some roadblocks in Hamilton's
program to prove the conjecture using the `Ricci flow'
\cite{hamilton1,caochow}.   It is therefore important to understand in
detail the properties of this flow.

In the following, we follow up on a suggestion by Bakas to use his 2D results
in order to analyze the flow equations for 3D manifolds with a single Killing vector.
This provides a tractable midisuperspace approach which can be systematically studied in the context of the `stringy flow' first 
considered in \cite{stringy}. We will show that this flow
reduces to the infinite dimensional generalization of the 
Toda equation for the conformal factor of the invariant 2D
submanifold plus a linear equation for the scale factor of the extra dimension.
 Note that since the latter scale factor depends on the coordinates of the invariant
 subspace, our manifolds are not simple direct products. In addition, we will analyze two
 exact analytic solutions in detail and show that they have the expected behaviour.

The paper is organized as follows. Section 2 reviews 2D flow equations and
Bakas' results, Section 3 reviews the stringy flow of \cite{stringy}, but with 
the De Turck modification \cite{deturck}.  The De Turck modification contains 
a vector field $\xi_i$, and we show that if we choose this vector field as a 
linear combination of two vector fields, one of which is proportional to the 
gradient of the dilaton field, then the dilaton can be decoupled from the flow of 
the remaining fields.  In Section 4 we discuss the flow for a metric ansatz where 
one of the coordinates is in the direction of a Killing vector field, and the 
remaining part of the metric is in the form of a conformal 2D metric. We use the second part of the De Turck vector field to preserve this form of
the metric throughout the flow.   In order 
that 
the flow is self-consistent, the U(1) vector field in the stringy flow must be 
fixed in terms of the functions that occur in the metric tensor.  The flow then 
reduces to two equations for the two metric degrees of freedom.  One of these is 
the `continual Toda equation' \cite{bakas} for the conformal factor of the 2D geometry 
orthogonal to the Killing vector field, and the other, for the component of the metric 
in the direction of the Killing vector field, is the linearization 
of the continual Toda equation.
Section 5 presents specific
solutions and Section 6 ends with conclusions and prospects for future work.

\section{The 2D Case}

We now summarize the methodology and results of Bakas\cite{bakas} since they play a crucial
role in the following. The Ricci flow equations, for arbirary 2-metric
$g_{AB}$ are:
\be
{\partial g_{ij}\over \partial t} = -R_{ij} + \nabla_i \xi_j + \nabla_j \xi_i.
\label{2D flow}
\ee
The last two terms (the so-called ``De Turck'' terms) incorporate the effects of all possible diffeomorphisms and can be chosen 
arbitrarily in order to simplify the equations and/or optimize convergence. 
Originally, De Turck \cite{deturck} chose the vector field 
\be
\xi^i:=g^{jk}\left(\Gamma^i_{jk}-\Delta^i_{jk}\right),
\ee
where $\Gamma^i_{jk}$ is the Christoffel connection with respect to the 
Riemannian metric $g_{ij}$ and $\Delta^i_{jk}$ is a fixed 
`background connection'.  The purpose was to replace the Ricci flow, which 
is only weakly parabolic, by an equivalent flow which is strongly parabolic.

Bakas chose to work in the conformal gauge:
\be
ds^2 = g_{ij}dx^idx^j= {1\over2}\exp(\Phi)(dx^2+dy^2).
\ee
In this gauge there is no need to add De Turck terms and the flow takes the form of a non-linear ``heat equation'':
\be
{\partial\over\partial t}e^\Phi = \nabla^2\Phi.
\label{2D heat equation}
\ee
 
The Toda equations describe the integrable interactions of a collection of two dimensional
fields $\Phi_i(x,y)$ coupled via the Cartan matrix $K_{ij}$:
\be
\sum_jK_{ij}e^{\Phi_j(x,y)}= \nabla^2\Phi_i(x,y).
\ee 
Bakas argues that Eq.(\ref{2D heat equation}) is a continual analogue of the above,
with the Cartan matrix replaced by the kernel:
\be
K_{ij}\to K(t,t')={\partial\over\partial t}\delta(t,t').
\ee
This leads to a general solution to (\ref{2D heat equation}) in terms of a power series
around the free field expanded in path ordered exponentials. Although the resulting expression
is difficult to work with explicitly, it does provide a formal complete solution to the 2D flow
equations.

In the next sections we will show that  a similar formal solution can also be found for 
three dimensional metrics with at least one Killing field.

\section{3D Flow Equations}

We consider here a generalization of the Ricci flow, in which, besides the metric $g_{ij}$, there are additional fields which flow, consisting 
of a dilaton $\phi$, a gauge two-form potential $B_{ij}$ with  
field strength $H_{ijk}$ and finally, a $U(1)$ gauge field 
with potential 1-form $A_i$ and corrresponding field strength $F_{ij}$ which couples as a
`Maxwell-Chern-Simons theory'.
Including De Turck terms plus gauge terms for the flow of the non-metric fields, the flow is
\bea
{\dot g}_{ij}&=&-2\left(R_{ij}+2\phi_{|ij}-(\epsilon_FF_i{}^k F_{jk}+\frac{\epsilon_H}{4} H_{ikl}H_j{}^{kl})\right)
+L_\xi g_{ij}, \label{gflow}\\
{\dot A}_i&=&-\left(e^{2\phi}\nabla_j(e^{-2\phi}F_i{}^j)+\frac{e\epsilon_F}{2}\eta_i{}^{jk}F_{jk}\right)
+L_\xi A_i + \partial_i\Lambda,\label{Aflow}\\
{\dot B}_{ij}&=&e^{2\phi}\nabla_k(e^{-2\phi}H^k{}_{ij})+L_\xi B_{ij} +\partial_i\Lambda_j - \partial_j\Lambda_i,\label{Bflow}\\
\dot\phi&=&-\chi +\Delta\phi-|\nabla\phi|^2+\frac{\epsilon_F}{2}F^2+\frac{\epsilon_H}{12}+L_\xi\phi.
\label{dilflow}
\eea
In the above,  $L_\xi$ denotes the Lie 
derivative with respect to the (covariant) vector field $\xi_i$. The Lie derivative terms are present because we are flowing geometrical objects that are not coordinate invariant, so their time derivatives should only be determined up to arbitrary gauge transformations at each point along the flow. Similarly, the terms containing $\Lambda$ and $\Lambda_i$ correspond to arbitrary gauge transformations on the gauge fields. By choosing the gauge and coordinate transformation terms judiciously, we are able to simplify the equations considerably.

This flow is motivated by two considerations.  First, 
as shown in \cite{stringy}, all of 
the Thurston geometries are solutions 
of the equations of motion of this theory for various values of 
the parameters $\chi,\epsilon_H,\epsilon_F,e$, as well as the other fields. 
In particular, the addition of the Maxwell term alone ($e=0$) yields 
$S^2\times E^1$, $H^2\times E^1$ and $Sol$ as solutions. Moreover, there exists a generalized Birkhoff theorem which guarantees that these are the only solutions when $\phi=constant$ and $A\neq0$. With $e\neq0$, one finds that the remaining Thurston geometries $Nil$ and $SL(2,R)$ are also solutions. As argued in \cite{stringy} it seems plausible that these are the only solutions, but to date no rigorous proof exists.

The second motivation comes from string theory.  In particular, the RG flow for a non-linear sigma model with a 4D Kaluza-Klein target space resembles the flow 
above, with the $A_i$ potential originating as the twist potential of the 4D Kaluza-Klein metric.  The details of this are being investigated elsewhere 
\cite{gegsun}.

We choose $\xi_i=k_i+2\nabla_i\phi$ and let $\Lambda=-2A_j\phi^{|j}$ and
$ \Lambda_i = 2B_{ji}\phi^{|j}$, where $k_i$ is as yet arbitrary.  With these choices the dilaton is completely eliminated from the flow equations for the metric and gauge fields:
\bea
\dot g_{ij}&=&-2\left(R_{ij}-(\epsilon_FF_i{}^k F_{jk}+\frac{\epsilon_H}{4} H_{ikl}H_j{}^{kl})\right)
+L_k g_{ij}, \label{gflowt}\\
\dot A_i&=&-\left(\nabla_jF_i{}^j+\frac{e\epsilon_F}{2}\eta_i{}^{jk}F_{jk}\right)
+L_k A_i+\partial_i \lambda,\label{Aflowt}\\
\dot B_{ij}&=&\nabla_k H^k{}_{ij}+L_k B_{ij}+\partial_i\lambda_j-\partial_j \lambda_i,\label{Bflowt}\\
\dot\phi&=&-\chi+\Delta\phi-2 |\nabla\phi|^2+\frac{\epsilon_F}{2}F^2+\frac{\epsilon_H}{6}H^2+L_k\phi.
\label{dilflowt}
\eea
The arbitrary vector $k^i$ and gauge parameters $\lambda,\lambda_i$ indicate that we are still free to add further De Turck 
and gauge terms to the equations. We will use this freedom later to simplify the equations that result from a particular {\it ansatz}.

\section{A Particular Case with One Killing Vector Field}

Henceforth we set $B_{ij}$ identically equal to $0$, which is consistent with the flow equations. We also consider the case $e=0$ (no Chern-Simons term).  We assume the metric to have a single Killing vector and to be manifestly hypersurface orthogonal (i.e. diagonal):
\be
ds^2=e^\Phi(dx^2+dy^2)+e^\sigma dw^2,
\ee
We also choose the following {\it ansatz} for the vector potential:
\be
A_i=[e^{\sigma/2},0,0].
\ee

Consistency of the above {\it ansatz} requires that the flow equations preserve the diagonal nature of the metric. It turns out that this can be accomplished by 
choosing the vector field $k_i$ as
\be
k_i=-{1\over2}\partial_i\sigma,
\ee

With these choices the flow equations simplify to:

\bea
\dot{g}_{xx}=e^\Phi\dot{\Phi}&=& \nabla^2 \Phi + {1\over2}(1+\epsilon_F)(\partial_x \sigma)^2,
                   \nonumber\\
\dot{g}_{yy}=e^\Phi\dot{\Phi}&=& \nabla^2 \Phi + {1\over2}(1+\epsilon_F)(\partial_y \sigma)^2,
                   \nonumber\\
                    \nonumber\\
\dot{g}_{xy}=0&=& {1\over2}(1+\epsilon_F) \partial_x\sigma\partial_y\sigma,
		     \nonumber\\
\eea
\bea
\dot{A}_x &=& \dot{A}_y = 0,\nonumber\\
\dot{A}_w&=&\epsilon_f \nabla^2 e^{\sigma/2} + \left({1+\epsilon_F}\right)
       \left(\partial\sigma\right)^2.
\eea
In the above, $\nabla^2$ denotes the flat space Laplacian.

We now fix $\epsilon_F=-1$, in which case the flow boils down to two simple partial differential equations.  The first is
\be
\partial_t e^\Phi=\nabla^2 \Phi,\label{ctoda}
\label{toda}
\ee
which is the `continual Toda eqn' a la  \cite{bakas}.  The other flow is the linearization 
of the continual Toda flow: 
\be
e^\Phi\partial_t e^{-\sigma/2}= -\nabla^2 e^{-\sigma/2}.\label{linToda}
\ee

To the best of our knowledge, there is no `simpler flow' constructed from the 
Ricci-De Turck flow alone, without other fields, which can self-consistently flow 
the metric preserving the manifestly static form. Note that for any given solution $\Phi$ of (\ref{toda}), there exists a corresponding solution for $\sigma$:
\be
e^{-\sigma/2} = \Phi(x,y,-t) + \chi(x,y), 
\ee
where $\chi(x,y)$ is any harmonic function on the $x,y$ subspace, i.e. satisfying: $\nabla^2\chi=0$.

\section{Exact Solutions of the Flow}

We first examine a non-trivial flow, namely the sausage solution of Bakas \cite{bakas} 
(also called the Rossineau flow in the mathematical literature \cite{caochow}).  This 
is an exact solution of the continual Toda equation of the form:  
\be
e^\Phi={2\sinh{[2\gamma t]}\over \gamma(\cosh{[2\gamma t]}+\cosh{(2 y)})}.
\ee
In this case:
\be
e^{-\sigma/2}= ln\left|{2\sinh{[2\gamma t]}\over \gamma(\cosh{[2\gamma t]}+\cosh{(2 y)})}\right|,
\ee
where we have eliminated an imaginary term from $e^{-\sigma/2}$ by using the freedom to shift by a harmonic function.

In the limit as $t\to\infty$, $e^\Phi\to 2/\gamma$, and $e^\sigma \to ln|2/\gamma|^{-2}$, so that the Ricci tensor 
goes to zero and in this limit, the geometry is flat.  On the other 
hand, in the limit
$t\to 0^+$, $e^\Phi\to 2t/\cosh^2{y}$.  In this limit, we find
that the Ricci scalar $R\sim\frac{1}{t}$.  
So, if we 
flow the highly curved non-homogeneous metric with initial value at $t=\epsilon>0$
\be
ds^2=\left(\ln{\frac{2\epsilon}{cosh^2{y}}}\right)^{-2}dw^2+\frac{2\epsilon}{\cosh^2(y)}\left(dx^2+dy^2\right),
\ee
we end up at $t\to\infty$ with the flat metric.  This is consistent with 
Thurston's conjecture.

The second type of solution is of the Liouville type.  We set
\be
e^{\Phi(x,y;t)}=T(t) e^{\psi(x,y)}.
\ee
Now for $t\geq 0$, we find that
\bea
T(t)&=&\beta t,\none
\nabla^2\psi&-&\beta e^\psi=0,
\eea
where $\beta$ is a separation constant.  The second of the above 
equations is the Liouville equation, so the two dimensional part of the metric,
$e^\Phi(dx^2+dy^2)$ has constant negative curvature (for $t\geq 0$).  

Again we choose $e^{-\sigma/2}=\Phi$, so that 
\be
e^\sigma=\left[\log {\beta t}+\psi(x,y)\right]^{-2}.
\ee
The separation constant $\log\beta $ can be absorbed into $\psi$ without loss of 
generality.

Hence the metric is
\be
ds^2=\left[\log{t}+\psi(x,y)\right]^{-2} dw^2+\beta t e^{\psi(x,y)}(dx^2+dy^2).\label{lioumetric}
\ee
The quantity $\psi(x,y)$ is a solution of the Liouvile equation.

If $t\geq 0$, then the flow starts from 
some highly curved non-homogeneous metric near $t=0$.  As $t\to \infty$, we have 
\bea
R_{AB}&\sim&-\frac{1}{2t}g_{AB},\none
R_{ww}&\sim& 0,
\eea
with $A,B,...=x,y$.  Hence, the geometry is asymptotically that of the 
homogeneous, but anisotropic geometry $H^2\times E^1$.

Thus the flow is consistent with the Thurston conjecture.  

\section{Conclusions}

We have shown that the modified Ricci flow equations Eq(9) for 3D metrics with at least one
Killing vector can be integrated in precisely the same manner as the 2D equations, at least for the special case $\epsilon_F=-1$. In addition to extending this analysis to other values of the parameters in the action, and hence to other topologies, it is interesting to speculate
whether these techniques could work for more general 3D metrics. 

Consider, without loss of generality, a diagonal metric
\be
ds^2=e^{\Phi_1(x;t)}(dx^1)^2+e^{\Phi_2(x;t)}(dx^2)^2+e^{\Phi_3(x;t)}(dx^3)^2,
\ee
where the functions $\Phi_i(x;t)$ depend on all 3 coordinates $x^i$.
The resulting  bare Ricci flow is again not manifestly elliptic, and the equations 
have non-trivial off-diagonal terms on the RHS that make direct integration difficult. Since in three
dimensions any metric can be made diagonal with a suitable coordinate transformation, it is
reasonable to assume that there exists a modified flow that ensures that diagonal metrics evolve into diagonal metrics. We have as yet not succeeded in constructing this modified flow, but if it did exist, it
is possible that the resulting three flow equations for each of the three scale factors
 would take a form similar to what we have found above, albeit with non-trivial coupling. It may therefore
provide a basis for solving the 3D flow equations in a more general setting.

\noindent
{\bf Acknowledgements}:  

We wish to thank the Perimeter Institute, where much 
of this work was done, for its 
hospitality and support. This work was supporte by the Natural Sciences and Engineering Research Council of Canada.
We would also like to thank Viqar Husain, Vardarajan Suneeta and Eric Woolgar for very useful discussions.

\end{document}